\documentclass[aps,prb,twocolumn,floatfix,showpacs,superscriptaddress]{revtex4-1}
\usepackage{graphicx,amsmath,amsfonts}

\begin{document}
\title{First-principles study of the magnetic ground state in kagome francisites Cu$_{3}$Bi(SeO$_{3}$)$_{2}$O$_{2}$X (X=Cl, Br)}
\author{S.~A.~Nikolaev}
\email{saishi@inbox.ru}
\affiliation{Department of Theoretical Physics and Applied Mathematics,
Ural Federal University, 620002 Yekaterinburg, Russia}
\author{V.~V.~Mazurenko}
\affiliation{Department of Theoretical Physics and Applied Mathematics,
Ural Federal University, 620002 Yekaterinburg, Russia}
\author{A.~A.~Tsirlin}
\affiliation{Experimentalphysik VI, Universit$\ddot{a}$t Augsburg, 83615 Augsburg, Germany}
\author{V.~G.~Mazurenko}
\affiliation{Department of Theoretical Physics and Applied Mathematics,
Ural Federal University, 620002 Yekaterinburg, Russia}
\date{\today}


\begin{abstract}
We explore magnetic behavior of kagome francisites Cu$_{3}$Bi(SeO$_{3}$)$_{2}$O$_{2}$X (X = Cl and Br) using first-principles calculations. To this end, we propose an approach based on the Hubbard model in the Wannier functions basis constructed on the level of local-density approximation (LDA). The ground-state spin configuration is determined by a Hartree-Fock solution of the Hubbard model both in zero magnetic field and in applied magnetic fields. Additionally, parameters of an effective spin Hamiltonian are obtained by taking into account the hybridization effects and spin-orbit coupling. We show that only the former approach, the Hartree-Fock solution of the Hubbard model, allows for a complete description of the anisotropic magnetization process. While our calculations confirm that the canted zero-field ground state arises from a competition between ferromagnetic nearest-neighbor and antiferromagnetic next-nearest-neighbor couplings in the kagome planes, weaker anisotropic terms are crucial for fixing spin directions and for the overall magnetization process. We thus show that the Hartree-Fock solution of an electronic Hamiltonian is a viable alternative to the analysis of effective spin Hamiltonians when a magnetic ground state and effects of external field are considered.
\end{abstract}

\maketitle

\section{Introduction}
\par The interest in novel frustrated magnets is continuously increasing because of the diversity of their crystal structures and magnetic configurations leading in turn to many interesting phenomena that may be relevant to applications, such as spintronics.\cite{landau,wold} For example, systems composed of $S=\frac{1}{2}$ spins on a two-dimensional kagome-type lattice are usually strongly frustrated due to competing exchange couplings that give rise to various exotic ground-states and peculiar properties at finite temperatures and in applied magnetic fields.\cite{balents}
\par Due to difficulties in synthesizing inorganic compounds with a low-dimensional arrangement of transition-metal cations, one usually resorts to special strategies for their construction. For instance, inorganic polyanionic groups can be used as a crystal structure spacer to form open volumes excluding bonding along one or two directions of the crystal structure. An ideal candidate for this purpose are the stereochemically active lone-pair cations (Se$^{+4}$ or Bi$^{+3}$), which are large enough to expand crystal structure and form bonds only with oxygen ions.\cite{galy,john1} Moreover, halogens are also often used, as they typically have low coordination numbers and additionally act as terminating species.\cite{john2} 
\par Cu$_{3}$Bi(SeO$_{3}$)$_{2}$O$_{2}$X with X = Cl, Br and I, hereinafter referred to as CBSOOX, are geometrically frustrated layered compounds with non-collinear magnetic ordering.\cite{millet} They can be derived from the natural mineral francisite and crystallize in the orthorhombic $Pmmn$ structure. This structure can be described as formed by copper(II)-oxygen layers with two nonequivalent copper sites. Cu1 and Cu2 of the site symmetries $-1$ and $mm2$, respectively, that build up a hexagonal network of CuO$_{4}$ square plaquettes linked together by Se$^{+4}$ and Bi$^{3+}$ ions with different out-of-plane oxygen bonding. This geometry can be regarded as a buckled kagome-type lattice.
\begin{figure*}
\includegraphics[scale=0.72]{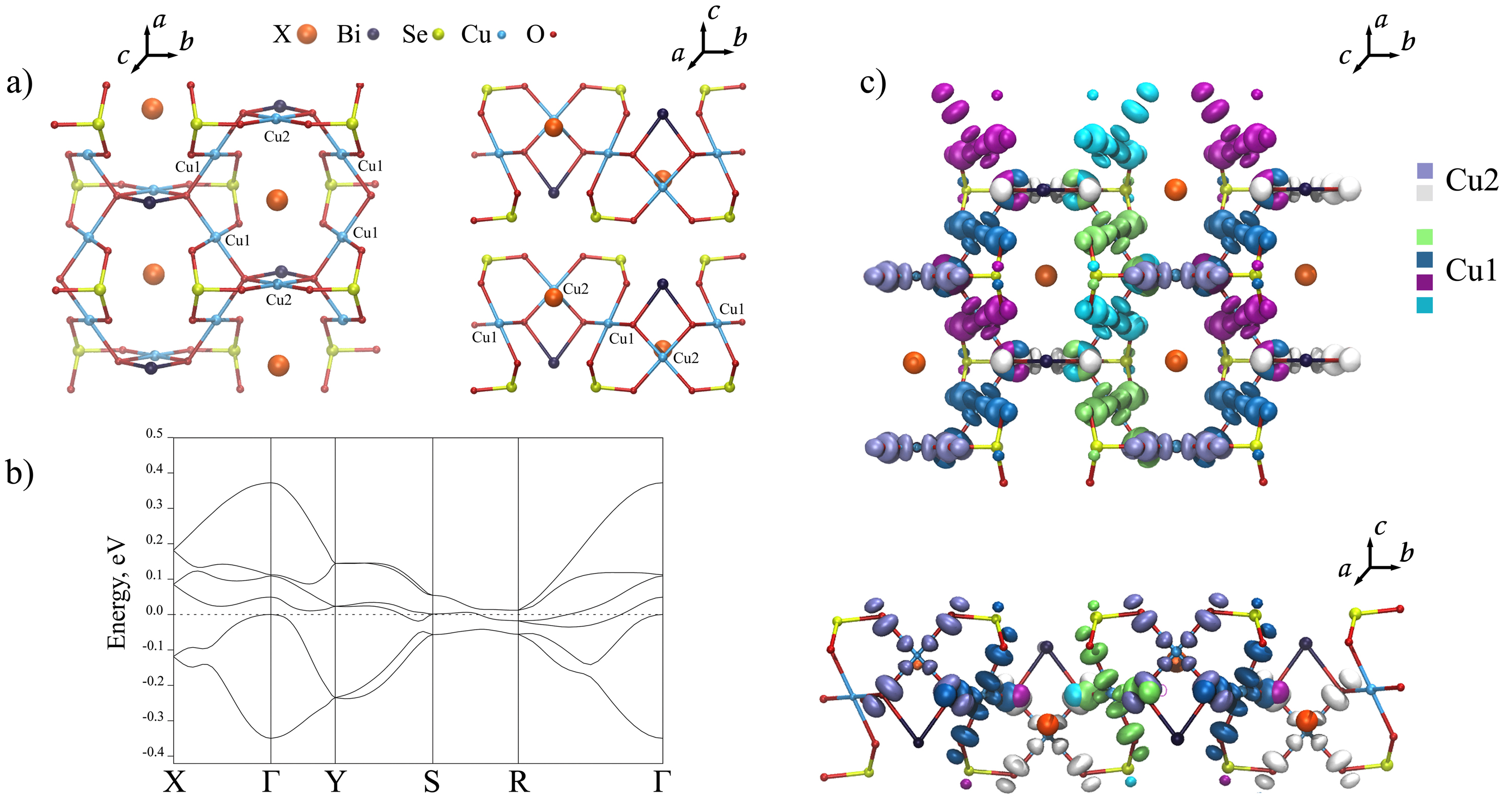}
\caption{a) Crystal structure of CBSOOX: $\emph{ab}$ (left) and $\emph{bc}$ (right) projections; b) Bands located near the Fermi level and corresponding to the copper $d^{x^{2}-y^{2}}$ states; c) $d^{x^{2}-y^{2}}$ Wannier functions centred on copper atoms as obtained from LDA calculations.}
\end{figure*}
\par Anisotropic magnetic properties of CBSOOX were studied by bulk magnetization measurements\cite{millet} and single-crystal neutron diffraction\cite{pregelj} that revealed the magnetic ordering transition at $T_{N}=24.7$~K. In zero field, \mbox{CBSOOBr} is ordered antiferromagnetically. Magnetic moments on the Cu2 sites point along the $\boldsymbol{c}$ axis, whereas those on the Cu1 sites are canted from $\boldsymbol{c}$ towards $\boldsymbol{b}$, thus producing the net magnetic moment within each layer in the $ab$ plane. These moments cancel out macroscopically because of the weakly antiferromagnetic interlayer coupling that can be overridden by a magnetic field $B_{C}\approx0.8$ T applied along the $c$ axis. External field triggers a metamagnetic transition, wherein magnetic moments in every second layer flip, and the net magnetic moment is formed macroscopically.\cite{pregelj,wang} Qualitatively similar behavior was observed for the Cl compound\cite{miller2012}, although no details of its magnetic structure were reported.
\par Rousochatzakis~\textit{et al.}\cite{teor} used density-functional calculations to derive isotropic and anisotropic exchange couplings in kagome francisites. They have shown that the canted spin order in the $ab$ plane originates from the competition between ferromagnetic nearest-neighbor ($J_1\simeq -70$\,K) and antiferromagnetic next-nearest-neighbor ($J_2\simeq 50$\,K) interactions. The leading Dzyaloshinsky-Moriya component along the $\boldsymbol{a}$ direction ($d_{1a}$ in the notation of Ref.~\onlinecite{teor}) restricts spins to the $bc$ plane. However, the alignment of the net moment with the $\boldsymbol{c}$ direction could not be reproduced, and details of the anisotropic magnetization process were not described quantitatively.
\par Recent studies revealed new unusual features of kagome francisites. For example, these compounds exhibit field- and temperature-dependent microwave absorption over a broad range of frequencies. This effect can be used for microwave filtering\cite{preg22} and calls for a quantitative microscopic description of the magnetization process. In the following, we will show how this description can be achieved. The novelty of our approach lies in the direct treatment of the electronic Hamiltonian, where all isotropic and anisotropic exchange couplings are included implicitly. We then use the same band structure to derive parameters of the effective spin Hamiltonian, and demonstrate salient differences between these two approaches. 

\section{Ground-state properties}
Electronic-structure calculations have been performed in terms of the conventional local-density approximation (LDA)\cite{kohn} and projected augmented waves (PAW) formalisms\cite{blochl}, as implemented in the Vienna ab-initio simulation package.\cite{kresse} Hopping parameters including the spin-orbit interaction have been calculated in the Quantum-ESPRESSO package.\cite{espr} In our calculations, we adopt the experimental crystallographic data obtained from single-crystal X-ray diffraction studies (Fig.~1a).\cite{millet} 
\par The LDA band structure of CBSOOX is presented in Fig.~1b. One of the most important features of the electronic structure is the presence of six bands located near the Fermi level and well separated from the rest of the spectrum. The main contribution to these bands comes from copper $3d^{x^{2}-y^{2}}$ states, which are strongly hybridized with oxygen $2p$ orbitals. There is also a small contribution coming from selenium and bismuth ions. To parametrize the LDA results, we have constructed the maximally-localized Wannier functions centered on the copper sites by projecting the Bloch states located near the Fermi level onto $d^{x^{2}-y^{2}}$ orbitals.\cite{wannier,most} As one can see from Fig.~1c, every two Wannier functions overlap only on one oxygen site (O1) giving rise to the so-called Coulomb contribution to the total exchange.\cite{maz1} The resulting functions have a noticeable contribution from both oxygen and selenium atomic orbitals, so they can be roughly expressed as the following linear combinations:
\begin{equation}
W_{1}=\alpha_{1}\phi_{Cu1}^{x^{2}-y^{2}}+2\beta_{1}\phi_{O1}^{p}+2\beta_{2}\phi_{O2}^{p}+2\gamma\phi_{Se}
\end{equation}
\noindent and
\begin{equation}
W_{2}=\alpha_{2}\phi_{Cu2}^{x^{2}-y^{2}}+2\beta_{1}\phi_{O1}^{p}+2\beta_{3}\phi_{O3}^{p}
\end{equation}
\noindent for Cu1 and Cu2 types, respectively, where contributions from bismuth and halogen orbitals are supposed to be negligible. These expansion coefficients for the Wannier functions can be estimated by using magnetic moments obtained from LSDA+U calculations.

\begin{table} 
\caption{Magnetic moments (in $\mu_{B}$) of Cu$_{3}$Bi(SeO$_{3}$)$_{2}$O$_{2}$X obtained from the LSDA+U calculations for the ferromagnetic configuration.}
\label{magmom}
\begin{tabular}{c|cccccccc}
\hline
\hline
&&&&&&&\vspace{-0.45cm}\\
X& $m_{\textrm{Cu1}}$& $m_{\textrm{Cu2}}$& $m_{\textrm{O1}}$& $m_{\textrm{O2}}$& $m_{\textrm{O3}}$& $m_{\textrm{Se}}$& $m_{\textrm{Bi}}$\vspace{-0.3cm}\\
&&&&&&&\\
\hline
Cl&0.757&0.746&0.160&0.046&0.047&0.018&0.006\vspace{-0.3cm}\\
&&&&&&&\\
Br&0.758&0.750&0.158&0.044&0.042&0.020&0.006\\
\hline
\hline
\end{tabular}
\end{table}

\par To reproduce the insulating ground states of the CBSOOX compounds, we performed LSDA+U calculations. In these calculations the value of the on-site Coulomb interaction and intra-atomic exchange interaction are chosen to be 9 eV and 1 eV, respectively. We have obtained the energy gap of 2.9 eV, which is close to the typical values for low-dimensional copper oxides. 
The calculated atomic magnetic moments used to estimate the expansion coefficients for the Wannier functions are presented in Table \ref{magmom}  and also indicate a strong metal-ligand hybridization. 

\section{Construction of the Hubbard model}
To describe electronic and magnetic properties of CBSOOX, we construct the following one-orbital Hubbard model:\cite{hubb}
\begin{eqnarray}
\hat{\cal H}&=&\sum_{\boldsymbol{R}\boldsymbol{R'},\sigma\sigma'}t_{\boldsymbol{R}\boldsymbol{R'}}^{\sigma\sigma'}\hat{a}_{\boldsymbol{R}\sigma}^{+}\hat{a}_{\boldsymbol{R'}\sigma'}\nonumber\\
&+&\frac{1}{2}\sum_{\boldsymbol{R},\sigma\sigma'}U_{\boldsymbol{R}}\,\hat{a}_{\boldsymbol{R}\sigma}^{+}\hat{a}_{\boldsymbol{R}\sigma'}^{+}\hat{a}^{}_{\boldsymbol{R}\sigma'}\hat{a}^{}_{\boldsymbol{R}\sigma}\\
&+&\frac{1}{2}\sum_{\boldsymbol{R} \boldsymbol{R'},\sigma\sigma'}J^{H}_{\boldsymbol{R} \boldsymbol{R'}}\,\hat{a}_{\boldsymbol{R}\sigma}^{+}\hat{a}_{\boldsymbol{R'}\sigma'}^{+}\hat{a}^{}_{\boldsymbol{R}\sigma'}\hat{a}^{}_{\boldsymbol{R'}\sigma}\nonumber,
\end{eqnarray}
\noindent where $\hat{a}_{\boldsymbol{R}\sigma}^{+}$ $(\hat{a}_{\boldsymbol{R}\sigma})$ creates (annihilates) an electron of spin $\sigma$ at site $\boldsymbol{R}$, $U_{\boldsymbol{R}}$ and $J^{H}_{\boldsymbol{R} \boldsymbol{R'}}$ are the local Coulomb and non-local exchange interactions, respectively, $t_{\boldsymbol{R}\boldsymbol{R'}}^{\sigma \sigma'}$ is the element of the hopping matrix with spin-orbit coupling. The transfer integrals are to be found from a
Wannier parametrization of the first-principle band structure (LDA+SO) including
spin-orbit coupling.\cite{most} The results are presented in Table III. One can see that the matrices between nearest neighbors contain large imaginary and non-diagonal elements that, as we will show below, are responsible for anisotropic exchange interactions. In turn, parameters $U_{\boldsymbol{R}}$ and  $J^{H}_{\boldsymbol{R} \boldsymbol{R'}}$ for the one-orbital model have been estimated by using the on-site Coulomb interaction for the $3d$ shell of the copper atom and intra-atomic exchange interaction of the ligand atoms as described in Ref. \onlinecite{maz1}.   
\par The constructed model Eq. (3) can be solved in the mean-field Hartree-Fock approximation, which gives us an opportunity to study field dependence of the magnetic ground state. Alternatively, one can evaluate both isotropic and anisotropic exchange interactions using second-order perturbation theory and thus obtain individual parameters of the effective spin Hamiltonian.

\section{Microscopic analysis of magnetic interactions} 
In the limit $t\ll U$, one can construct the Heisenberg-type model by using the superexchange theory proposed by Anderson:\cite{Anderson, moriya}
\begin{eqnarray}
H = \sum_{\boldsymbol{R} \boldsymbol{R'}} J_{\boldsymbol{R} \boldsymbol{R'}} \hat{ \boldsymbol{S}}_{\boldsymbol{R}} \hat{\boldsymbol{S}}_ {\boldsymbol{R'}}+ \sum_{\boldsymbol{R} \boldsymbol{R'}} \boldsymbol{D}_{\boldsymbol{R} \boldsymbol{R'}} [ \hat{ \boldsymbol{S}}_{\boldsymbol{R}} \times \hat{\boldsymbol{S}}_{\boldsymbol{R'}}], 
\label{heis}
\end{eqnarray}
where $J_{\boldsymbol{R} \boldsymbol{R'}}$ and $\boldsymbol{D}_{\boldsymbol{R} \boldsymbol{R'}}$ are the isotropic and anisotropic (Dzyaloshinsky-Moriya) exchange interactions between $\boldsymbol{R}$th and $\boldsymbol{R'}$th sites, respectively. 
\par In terms of the Hubbard model parameters Eq.~(3), the isotropic exchange interaction is expressed in the following form:\cite{Anderson, Aharony}
\begin{eqnarray}
J_{\boldsymbol{R} \boldsymbol{R'}} = \frac{2}{U_{\boldsymbol{R}}} {\rm Tr} \{ \hat t_{\boldsymbol{R'} \boldsymbol{R}} \hat t_{\boldsymbol{R} \boldsymbol{R'}} \} - 2 J^{H}_{\boldsymbol{R} \boldsymbol{R'}}
\label{exch}
\end{eqnarray}
\noindent The first term in Eq.(\ref{exch}) is the kinetic Anderson's exchange interaction taking into account spin-orbit coupling, while the second one is a ferromagnetic exchange coming from the overlap of the Wannier functions.\cite{maz1} As it is seen from Fig. 2, it has a non-zero value for the 1--4 and 1--5 pairs. Generally, $J^{H}_{\boldsymbol{R} \boldsymbol{R'}}$ in Eq.\ref{exch} can be estimated directly from LDA+SO calculations by integrating the corresponding combination of the Wannier functions. On the other hand, its value can be approximated as follows:
\begin{equation}
J^{H}_{\boldsymbol{R} \boldsymbol{R'}}=\beta^{4}J^{p,O}_{H}+\gamma^{4}J^{p,Se}_{H},
\label{eq:fm}
\end{equation}
\noindent where  $J^{p,O}_{H}$ and $J^{p, Se}_{H}$ are the intra-atomic exchange interactions of oxygen and selenium ions, respectively. The values of intra-atomic exchange interactions $J^{p,O}_{H}$ and $J^{p,Se}_{H}$ can be also estimated from LSDA+U calculations through the shift of band centers for the majority $C^{\uparrow}$ and minority $C^{\downarrow}$ spin components, $J^{p,I}_{H}=(C_{I}^{\uparrow}-C_{I}^{\downarrow})/M_{I}$, where $M_{I}$ is the magnetic moment on the corresponding site. The values obtained for $J^{p,O}_{H}$ and $J^{p,Se}_{H}$ are 1.50 eV and 0.9 eV, respectively. 

\begin{figure}
\centering
\includegraphics[scale=0.3]{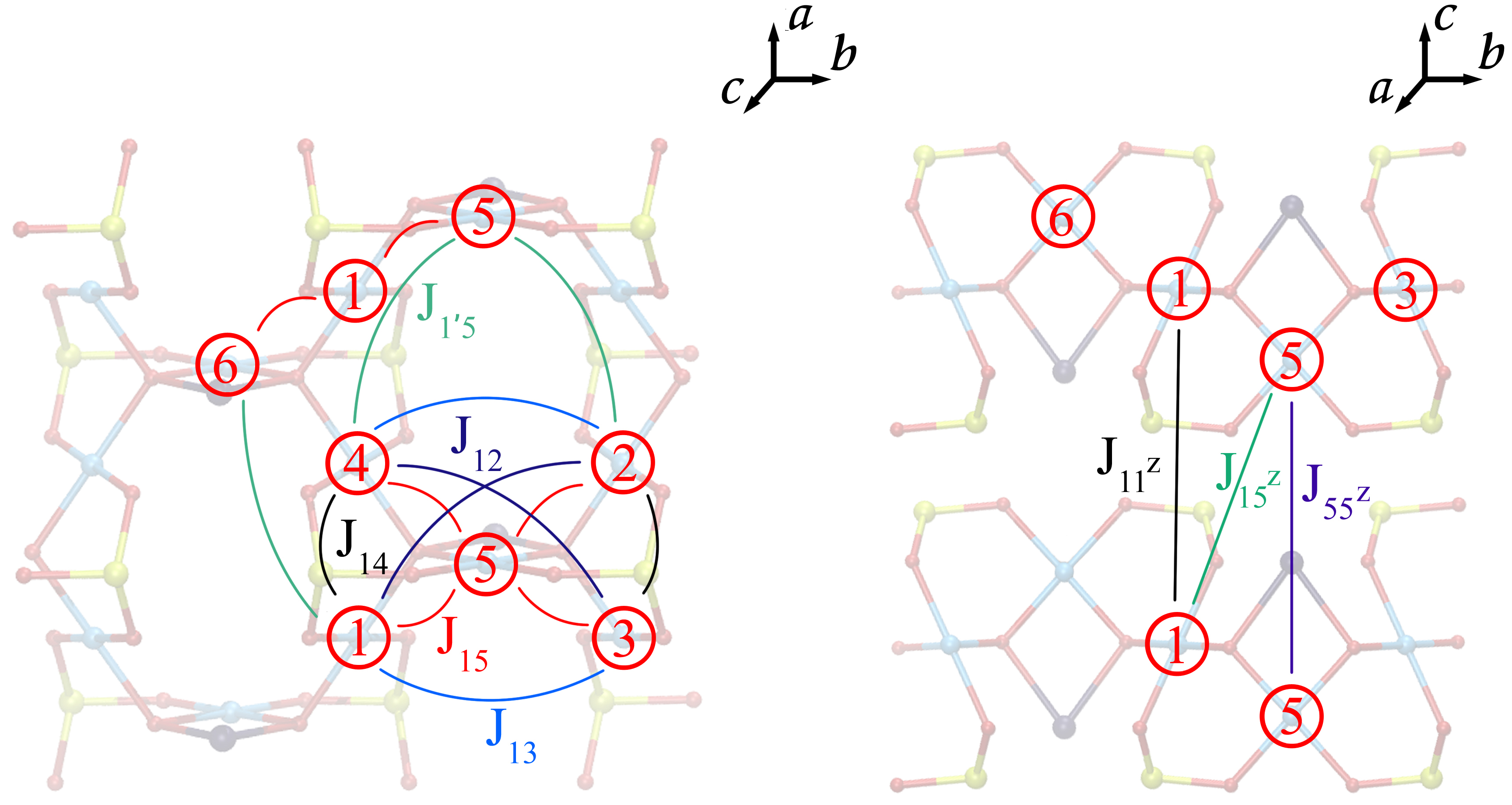}
\caption{Schematic view of intraplane (left panel) and interplane (right panel) isotropic exchange parameters. Sites 1, 2, 3, 4 and 5, 6 stand for Cu1 and Cu2 atoms, respectively.}
\end{figure}

\par Hopping integrals between $d^{x^{2}-y^{2}}$ orbitals of the copper atoms obtained from LDA+SO calculations and the corresponding isotropic exchange parameters calculated from Eq. (\ref{exch}) are presented in Table II. We find a strong intraplane FM coupling defined by the nearest-neighbor Cu1--Cu2 and Cu1--Cu1 interactions, $J_{15}$ and $J_{14}$, respectively. The dominant intraplane AFM exchange interaction $J_{13}$ is found between the neighboring Cu1 sites, for which the corresponding magnetic orbitals do not overlap on the oxygen atoms. This coupling is about two times smaller than the FM one. Our results are in good agreement with the previous theoretical analysis\cite{teor} that arrived at the values of $J_{13}=4.74$ (5.08)~meV, $J_{14}=-6.54$ ($-6.46$)~meV and $J_{15}=-5.68$ ($-5.77$) for X = Cl (Br). However, they are rather different from the experimentally predicted ratio of $\approx$1.6, where the AFM interaction dominates.\cite{pregelj} The copper atoms in the adjacent layers are coupled antiferromagnetically, and the interplane interaction is essentially weak (about 10--20 times smaller than the corresponding intraplane interactions). The isotropic exchange parameters are very similar for X=Cl or Br, that implies a small contribution of halogen's $p$--states to the localized bands near the Fermi level. 

{\it Dzyaloshinskii-Moriya (DM) interactions.}
Anisotropic exchange parameters can be derived by extending the theory of superexchange interaction in the case of spin-orbit coupling. They have the following form:\cite{moriya, Aharony}
\begin{equation}
\label{anisanis}
\boldsymbol{D}_{\boldsymbol{R} \boldsymbol{R'}}=-\frac{i}{U_{\boldsymbol{R}}} [{\rm Tr} \{ \hat t_{\boldsymbol{R'} \boldsymbol{R}} \} {\rm Tr} \{ \hat t_{\boldsymbol{R} \boldsymbol{R'}}  \boldsymbol{\sigma} \}\!-\!{\rm Tr} \{ \hat t_{\boldsymbol{R} \boldsymbol{R'}} \} {\rm Tr} \{ \hat t_{\boldsymbol{R'} \boldsymbol{R}} \boldsymbol{\sigma} \}],
\end{equation}
\noindent where $\boldsymbol{\sigma}$ is a vector of the Pauli matrices. The values of anisotropic exchange parameters are presented in Table III. The dominant anisotropic exchange coupling is found between the nearest neighbors Cu1 and Cu2, and its component along the $\boldsymbol{a}$ axis is by far the largest among the DM couplings. The value of $\boldsymbol{D}_{15}$ is mostly determined by the relatively large hopping parameters $t_{15}$. Moreover, the $\boldsymbol{a}$ component of $\boldsymbol{D}_{15}$ has different orientations in two adjacent hexagonal networks of the same $ab$ layer, and is responsible for the stabilization of the canted magnetic order.\cite{teor} Our results for the exchange parameters obtained within Moriya's microscopic theory are in good agreement with the ones estimated from LDA+U+SO total energies.\cite{teor}

\begin{table}
\caption{Hopping integrals $t_{ij}$ (in meV) and corresponding isotropic exchange parameters $J_{ij}$  (in meV) between copper $d^{x^{2}-y^{2}}$ orbitals as obtained from LDA+SO calculations. Here, normal, dashed symbols and symbols with superscript $x$, $y$ and $z$ denote nearest neighbors, next-nearest neighbors and neighbors along the corresponding axis, respectively. Sites 1, 2, 3, 4 and 5, 6 represent Cu1 and Cu2 atoms, respectively. See Fig. 2 for details.}
\label{hoppings}
\begin{center}
\begin{tabular}{c|rr||c|rr}
\hline
\hline
&Cl&Br&&Cl&Br\\
&&&&&\\
\hline
Intraplane&&&&&\\
$\boldsymbol{t_{14}}$&\textbf{--1.2}&\textbf{-10.2}&$\boldsymbol{J_{14}}$&\textbf{--8.50}&\textbf{--8.29}\\
$\boldsymbol{t_{15}}$&\textbf{--45.8}&\textbf{--32.7}&$\boldsymbol{J_{15}}$&\textbf{--6.81}&\textbf{--7.43}\\
$\boldsymbol{t_{13}}$&\textbf{--78.4}&\textbf{--71.3}&$\boldsymbol{J_{13}}$&\textbf{4.77}&\textbf{4.02}\\
$t_{12}$&26.4&28.1&$J_{12}$&0.55&0.62\\
$t_{11^{x}}$&$-5.9$&$-11.4$&$J_{11^{x}}$&0.03&0.10\\
$t_{55^{y}}$&23.5&30.6&$J_{55^{y}}$&0.43&0.73\\
$t_{1'5}$&$-35.2$&$-31.5$&$J_{1'5}$&0.97&0.78\\
Interplane&&&&&\\
$t_{11^{z}}$&10.8&7.5&$J_{11^{z}}$&0.09&0.04\\
$t_{55^{z}}$&22.5&17.3&$J_{55^{z}}$&0.40&0.23\\
$t_{15^{z}}$&9.3&8.7&$J_{15^{z}}$&0.07&0.06\\
\hline
\hline
\end{tabular}
\end{center}
\end{table}

\begin{table*}
\caption{Hopping integrals $t_{ij}^{\sigma\sigma'}$ (in meV) and corresponding anisotropic exchange parameters $\boldsymbol{D}_{ij}$ (in meV) between copper $d^{x^{2}-y^{2}}$ orbitals for the one-orbital model as obtained from LDA calculations. See Fig. 2 for details.\vspace{0.2cm}}
\label{anisotropy}
\begin{tabular}{c|c|c}
\hline
\hline
&Cl&Br\\
\hline
&&\vspace{-0.3cm}\\
$\boldsymbol{t}_{14}$& $ \left(\begin{array}{cc}
-1.2- 5.6 i & 9.3 \\
-9.3           &-1.2 + 5.6 i \end{array}\right)$ & $\left(\begin{array}{cc}
-10.2 - 19.9 i & 17.6 \\
-17.6             &-10.2 +19.9 i  \end{array}\right)$  \\
&&\vspace{-0.3cm}\\
&&\vspace{-0.2cm}\\
$\boldsymbol{t_{15}}$& $\left(\begin{array}{cc}
-45.8 -12.9 i & 10.1- 25.1i \\
-10.1 -25.1 i & -45.8 + 12.9 i \end{array}\right)$ & $\left(\begin{array}{cc}
-32.7-8.3 i    &15.8 - 32 i \\
-15.8-32 i     &-32.7 + 8.3 i \end{array} \right)$\\
&&\vspace{-0.1cm}\\
$\boldsymbol{t_{13}}$&$\left(\begin{array}{cc}
-78.4-3.9 i  & -2.6 i \\
-2.6 i           &-78.4+3.9 i  \end{array} \right)$&$\left(\begin{array}{cc}
-71.3 - 2.4 i &-1.1 i \\
-1.1i             & -71.3+2.4 i  \end{array} \right)$\\
&&\vspace{-0.3cm}\\
\hline
&&\vspace{-0.35cm}\\
$\boldsymbol{D_{14}}$&$\mathbf{(0.0,\,0.10,\,0.06)}$&$\mathbf{(0.0,\,0.16,\,0.18)}$\\
$\boldsymbol{D_{13}}$&$\mathbf{(0.18,\,0.0,\,0.27)}$&$\mathbf{(0.07,\,0.0,\,0.15)}$\\
$\boldsymbol{D_{15}}$&$\mathbf{(1.02,\,0.41,\,0.33)}$&$\mathbf{(0.93,\,0.46,\,0.24)}$\\
&&\vspace{-0.35cm}\\
\hline
\hline
\end{tabular}
\end{table*}

\section{Hartree-Fock solution of the electronic Hamiltonian}
\par To study the influence of external magnetic field $\boldsymbol{B}$ on the magnetic ground state, we include the Zeeman interaction in the effective Hubbard model: 
\begin{equation}
\hat{\mathcal{H}}^{Z}=\sum_{\boldsymbol{R}}\mu_{B}g_{s}\,\boldsymbol{B}\cdot\hat{\boldsymbol{S}}_{\boldsymbol{R}},
\end{equation}
\noindent where $\hat{\boldsymbol{S}}_{\boldsymbol{R}}=\frac{1}{2}\sum_{\sigma\sigma'}\hat{a}^{}_{\boldsymbol{R}\sigma}\boldsymbol{\sigma}^{\sigma\sigma'}\hat{a}^{+}_{\boldsymbol{R}\sigma'}$ stands for the localized magnetic moment, $\mu_{B}$ is the Bohr magneton, and $g_{s}\approx2.0$ is the gyromagnetic ratio. Generally, the Zeeman term must include the full magnetic moment interacting with an external field. However, since the one-orbital model has been employed, the orbital magnetic moment is neglected and we can consider only its spin counterpart. By using the mean-field Hartree-Fock approximation, the one-orbital Hubbard model can be solved as follows:\cite{solhart}
\begin{equation}
\left(\hat{t}_{\mathbf{k}}+\hat{\mathcal{U}} +\hat{\mathcal{V}}^{H} + \hat{\mathcal{V}}_{\mathbf{k}}^{H}  \right)|\varphi_{\mathbf{k}}\rangle=\varepsilon_{\mathbf{k}}|\varphi_{\mathbf{k}}\rangle,
\end{equation}
\noindent where $\hat{t}_{\mathbf{k}}=\sum_{\boldsymbol{R}\boldsymbol{R'}}\hat{t}_{\boldsymbol{R}\boldsymbol{R'}}e^{-i\mathbf{k}(\boldsymbol{R}-\boldsymbol{R'})}$ is the Fourier image of hopping parameters with spin-orbit coupling, $|\varphi_{\mathbf{k}}\rangle$ are eigenvectors in the basis of Wannier functions, $\hat{\mathcal{U}}=\sum_{\boldsymbol{R}}\hat{\mathcal{U}}_{\boldsymbol{R}}$, $\hat{\mathcal{V}}^{H}=\sum_{\boldsymbol{R}}\hat{\mathcal{V}}^{H}_{\boldsymbol{R}}$ and $\hat{\mathcal{V}}^{H}_{\boldsymbol{k}}=\sum^{}_{\boldsymbol{R}\boldsymbol{R'}}\hat{\mathcal{V}}^{H}_{\boldsymbol{k},\boldsymbol{R}\boldsymbol{R'}}$ are the Hartree-Fock potentials, where the first term includes the on-site Coulomb and Zeeman interactions:
\begin{equation}
\hat{\mathcal{U}} _{\boldsymbol{R}}=\left( \begin{array}{cc}
\!\!\!U_{\boldsymbol{R}}n_{\boldsymbol{R}}^{\downarrow\downarrow}+\mu_{B}B_{z} &\!\!\!\!\!\!\!\!\!-U_{\boldsymbol{R}}n_{\boldsymbol{R}}^{\downarrow\uparrow}+\mu_{B}(B_{x}-iB_{y})  \\
\!\!\!-U_{\boldsymbol{R}}n_{\boldsymbol{R}}^{\uparrow\downarrow}+\mu_{B}(B_{x}+iB_{y})&\!\!\!\!\!\!\!\!\!U_{\boldsymbol{R}}n_{\boldsymbol{R}}^{\uparrow\uparrow}-\mu_{B}B_{z} \end{array}\!\! \right),
\end{equation}
\noindent while the other two terms originate from the non-local exchange coupling:
\begin{equation}
\hat{\mathcal{V}}^{H}_{\boldsymbol{R}}=-\sum_{\boldsymbol{R'}}J^{H}_{\boldsymbol{R}\boldsymbol{R'}}\left( \begin{array}{cc}
n^{\uparrow\uparrow}_{\boldsymbol{R'}}&n_{\boldsymbol{R'}}^{\downarrow\uparrow} \\
n_{\boldsymbol{R'}}^{\uparrow\downarrow}&n^{\downarrow\downarrow}_{\boldsymbol{R'}} \end{array} \right)
\end{equation}
\noindent and
\begin{equation}
\hat{\mathcal{V}}^{H} _{\mathbf{k},\boldsymbol{R}\boldsymbol{R'}}=J^{H}_{\boldsymbol{R}\boldsymbol{R'}}e^{i\mathbf{k}(\boldsymbol{R}-\boldsymbol{R'})}\!\left( \begin{array}{cc}
\!\!n^{\uparrow\uparrow}_{\boldsymbol{R}\boldsymbol{R'}}+n^{\downarrow\downarrow}_{\boldsymbol{R}\boldsymbol{R'}} & 0\\
0 &\!\!\!\!\!\!n^{\uparrow\uparrow}_{\boldsymbol{R}\boldsymbol{R'}}+n^{\downarrow\downarrow}_{\boldsymbol{R}\boldsymbol{R'}}
\end{array}\!\!\right),
\end{equation}
\noindent where the on-site and intrasite density matrices are defined as $\hat{n}_{\boldsymbol{R}} =\sum_{\boldsymbol{k}}|\varphi_{\mathbf{k}}\rangle\langle\varphi_{\mathbf{k}}|$ and $\hat{n}
_{\boldsymbol{R}\boldsymbol{R'}}  =\sum_{\boldsymbol{k}}|\varphi_{\mathbf{k}}\rangle\langle\varphi_{\mathbf{k}}|e^{-i\mathbf{k}(\boldsymbol{R}-\boldsymbol{R'})}$, respectively. The form of the full Hartree-Fock potential is in agreement with the Heisenberg model considered above, where the on-site Coulomb interaction $\hat{\mathcal{U}}$ is renormalized by the intra-atomic exchange $\hat{\mathcal{V}}^{H}$, while $\hat{\mathcal{V}}^{H}_{\boldsymbol{k}}$ represents ferromagnetic contribution to the total exchange.
\begin{figure}
\includegraphics[scale=0.42]{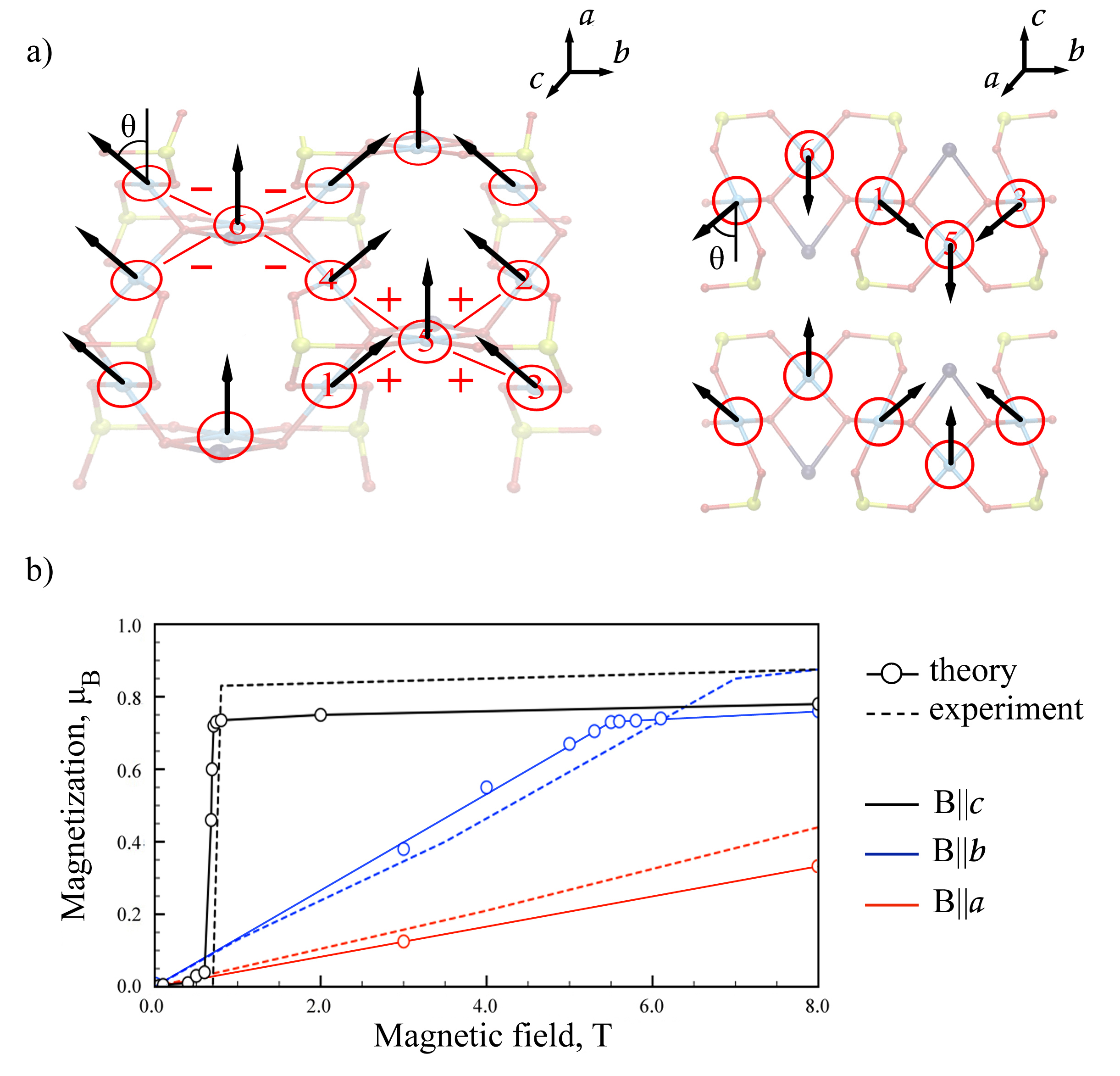}
\caption{a) Magnetic lattice as obtained from the mean-field Hartree-Fock approximation at zero applied magnetic field. Sites 1, 2, 3, 4 and 5, 6 stand for Cu1 and Cu2 atoms, respectively. The relative sign of the dominant $\boldsymbol{a}$ component of the anisotropic exchange coupling $\boldsymbol{D}_{15}$ is shown by $\pm$. b) Total magnetization per site in Cu$_{3}$Bi(SeO$_{3}$)$_{2}$O$_{2}$Br for three different directions of the magnetic field $\boldsymbol{B}$ and their comparison with experimental data.\cite{pregelj}}
\end{figure}
\par In the absence of external magnetic field, we find a canted order in the $ab$ planes and antiferromagnetic stacking of such ferrimagnetic layers (Fig. 3a). Magnetic moments at the Cu2 sites are strictly parallel to the $\boldsymbol{c}$ axis, while magnetic moments at the Cu1 sites alternate from this parallel direction towards the $\boldsymbol{b}$ axis and form a canted magnetic texture with the $bc$ angles $\theta=$ $50.1^{\circ}$ and $53.8^{\circ}$ for X = Cl and Br, respectively. This result is in good agreement with the experimental\cite{pregelj} and theoretical\cite{teor} values of the canting angle $\theta$ for X=Br, $51.6^{\circ}$ and $52.1^{\circ}$, respectively. The corresponding magnetic moments for X = Cl and Br are $\mathbf{m}_{\mathrm{Cu1}}=(0.01,\,0.77,\,0.64)\,\mu_{B}$ and $\mathbf{m}_{\mathrm{Cu2}}=(0.0,\,0.0,\,1.0)\,\mu_{B}$, $\mathbf{m}_{\mathrm{Cu1}}=(0.0,\,0.81,\,0.59)\,\mu_{B}$ and $\mathbf{m}_{\mathrm{Cu2}}=(0.0,\,0.0,\,1.0)\,\mu_{B}$, respectively. These directions are close to those obtained from the neutron diffraction measurements\cite{pregelj} for X=Br, $\mathbf{m}_{\mathrm{Cu1}}=(0.0,\,0.72,\,0.57)\,\mu_{B}$ and $\mathbf{m}_{\mathrm{Cu2}}=(0.0,\,0.0,\,0.90)\,\mu_{B}$. On the other hand, quantum renormalization of the magnetic moments (reduction from 1.0\,$\mu_B$ to $\sim 0.9$\,$\mu_B$ due to quantum fluctuations\cite{teor}) is not taken into account.

\begin{table*}
\caption{\label{tab:comparison} Experimental quantities and their predictions based on the LDA+SO, LDA+U+SO and Hartree-Fock (HF) calculations: in-plane canting angle in the $bc$ plane $\theta$, out-of-plane canting angle for Cu1 spins $\gamma$, slopes of the magnetization curves $k_{a}$ and $k_{c}$ for $\boldsymbol{B}\parallel \boldsymbol{a}$ and $\boldsymbol{B}\parallel \boldsymbol{c}$, respectively, and the work (per site) $W_{b}$ done by the magnetic field $\boldsymbol{B}\parallel \boldsymbol{b}$ going up to the characteristic field.
\vspace{0.2cm}}
\begin{tabular}{cl|ccccc}
\hline
\hline
& & $\theta$ & $\gamma$ & $\kappa_{a}$ & $\kappa_{c}$ & $W_{b}$ \\
& & (deg) & (deg) & ($\mu_{B}$/T) & ($\mu_{B}$/T) & ($\mu_{B}$T)\\
\hline
Br & LDA+SO & 38.5 & 1.3 & 0.24 & 0.0082 & 0.076 \\
& HF & 53.8 & 0.4 & 0.045 & 0.0077 & 2.0\\
& LDA+U+SO \cite{teor} & 52.1 & 0.7 & 0.16 & 0.0060 & 0.2 \\
& Experiment \cite{pregelj} & 51.6 & -- & 0.066 & 0.0074 & 2.9 \\
\hline
Cl & LDA+SO & 51.4 & 1.0 & 0.15 & 0.0073 & 0.046 \\ 
& HF & 50.1 & 0.8 & 0.044 & 0.0061 & 2.2 \\
\hline
\hline
\end{tabular}
\end{table*}
 
\par By applying symmetry operations of the $Pmmn$ space group, it follows that the magnetic ground state corresponds to the $\Gamma_{3}$ representation for both Cu sites. This representation allows for an additional canting out of the $bc$ plane, and, according to Ref.~\onlinecite{teor}, this canting should be indeed present in kagome francisites, although it has not been detected in the experiment so far. Indeed, our zero-field ground state derived from the Hartree-Fock solution of the electronic Hamiltonian features a weak out-of-plane canting $\gamma=0.8^{\circ}$ and $0.4^{\circ}$ for X = Cl and Br, respectively. The out-of-plane canting involves only the Cu1 spins and is much smaller than the in-plane canting. 
\par In the case of an external magnetic field $\boldsymbol{B}\parallel \boldsymbol{c}$, the zero-field AFM ground state is followed by a high-field ferrimagnetic state with the critical field $\boldsymbol{B}_{C}=0.87$ T and $0.74$ T for X = Cl and Br, respectively (Fig. 3b), in agreement with the experimental value of $\sim0.8$ T.\cite{pregelj}  Since the AFM components are still present, we do not observe a full saturation to the FM state. For the magnetic fields along the $\boldsymbol{a}$ and $\boldsymbol{b}$ axes, no metamagnetic transitions are observed, and the magnetic moments change continuously in each layer. In the case of $\boldsymbol{B}\parallel\boldsymbol{a}$, which acts against $\boldsymbol{D}_{15}$ and turns the magnetic moments out of the $bc$ plane, we get an almost linear behaviour of the magnetization up to $\sim 17.0$ T and $\sim 16.0$ T for X = Cl and Br, respectively, that is close to the (extrapolated) experimental value of $\sim15.0$ T.\cite{pregelj} For $\boldsymbol{B}\parallel\boldsymbol{b}$, which changes the direction of the net moment without flipping spins out of the $bc$ plane, the magnetization increases faster. Net moments of individual layers are polarized at $\sim 5.8$ T and $\sim 5.5$ T for X = Cl and Br, respectively, that is close to the experimental value of $\sim7.0$ T.\cite{pregelj} The smaller slope of the magnetization with respect to the magnetic field $\boldsymbol{B}\parallel\boldsymbol{a}$ is a result of the dominant $\boldsymbol{a}$ component of $\boldsymbol{D}_{15}$.

\section{Electronic vs. spin Hamiltonians}
\par We are now in a position to compare our Hartree-Fock solution of the electronic Hamiltonian to the effective spin Hamiltonian obtained from the \textit{same} LDA+SO band structure. To this end, we use several characteristic parameters derived in Ref.~\onlinecite{teor}. The primary canting angle $\theta$ and the out-of-plane canting angle $\gamma$ are given by:
\begin{equation}
J_{13}\sin{2\theta}+J_{15}\sin{\theta}+D_{15}^{a}\cos{\theta}-D_{13}^{a}\cos{2\theta}=0,
\end{equation}
\noindent and
\begin{equation}
\gamma=\frac{-D_{15}^{b}-D_{14}^{b}\cos{\theta}}{2J_{14}+J_{15}\,\sec{\theta}+D_{13}^{a}\tan{\theta}}.
\end{equation}
\noindent The slopes of the magnetization curves for $\boldsymbol{B}\parallel \boldsymbol{a}$ and $\boldsymbol{B}\parallel \boldsymbol{c}$ are given by $\kappa_a$ and $\kappa_c$, as follows: 
\begin{equation}
\kappa_{a}=\frac{(g\mu_{B})^{2}}{12}\frac{J_{15}(1+2\cos{\theta})^{2}\csc{\theta}+D_{13}^{a}-4D_{15}^{a}\cos{\theta}}{D_{15}^{a}(J_{15}+D_{13}^{a}\sin{\theta})-J_{15}D_{13}^{a}\cos{\theta}},
\end{equation}
\begin{equation}
\kappa_{c}=\frac{(g\mu_{B})^{2}/3}{2J_{13}+\left[D_{15}^{a}-D_{13}^{a}\cos{\theta}(1+2\sin^{2}{\theta})\right]/\sin^{3}{\theta}}.
\end{equation}
\noindent Finally, the slope of the magnetization curves for $\boldsymbol{B}\parallel \boldsymbol{b}$ is defined by the work $W_b$ required for polarizing spins in the field: 
\begin{equation}
W_{b}=\frac{1}{3}\frac{(D_{15}^{b}+D_{14}^{b}\cos{\theta})^{2}}{-2J_{14}+2J_{13}+D_{15}^{a}\csc{\theta}-D_{13}^{a}\cot{\theta}}.
\end{equation}
\noindent In Table~\ref{tab:comparison}, we also list experimental values along with the results of Ref.~\onlinecite{teor}, where parameters of the spin Hamiltonian are obtained from LDA+U+SO calculations.
\par The values based on the spin Hamiltonian parametrized via LDA+SO or LDA+U+SO calculations are qualitatively very similar, while minor quantitative differences can be traced back to different band-structure codes. The most tangible discrepancy is seen for the primary canting angle $\theta$, where LDA+SO underestimates the canting because of the overestimated ferromagnetic couplings $J_{14}$ and $J_{15}$. This overestimate is likely related to the approximate nature of Eq.~(\ref{eq:fm}) for non-90$^{\circ}$ ferromagnetic superexchange.
\par The Hartree-Fock solution of the electronic Hamiltonian mitigates the problem of the ferromagnetic superexchange and results in the realistic value of $\theta$. Even more importantly, this solution largely improves the description of the magnetization curves, where, for example, LDA+SO and LDA+U+SO underestimate $W_b$ by an order of magnitude. The Hartree-Fock solution correctly puts spins on the Cu2 atoms (and thus the net moment) along the $\boldsymbol{c}$ axis, which is not expected from the spin Hamiltonian, where both $b$- and $c$-components of $D_{15}$ ($d_{1b}$ and $d_{1c}$ in the notation of Ref.~\onlinecite{teor}) are clearly non-negligible suggesting a tilted direction of the net moment in the $bc$ plane.

\section{Summary and Conclusions}
\par We have shown that the spin Hamiltonian approach has its limitations for description of the magnetic anisotropy in kagome francisites. Deficiencies of the spin Hamiltonian are related to the fact that only the leading isotropic and anisotropic exchange couplings were included in the theoretical analysis of Ref.~\onlinecite{teor}. For example, symmetric components of the anisotropy, albeit weak, can also affect spin directions and peculiarities of the magnetization process, but in a complex system like francisite the inclusion of the symmetric anisotropy renders the spin Hamiltonian cumbersome and makes the whole problem intractable for analytical or numerical solution. In this case, the Hartree-Fock solution of the electronic Hamiltonian provides a viable alternative that delivers ground-state magnetic configuration both in zero and applied magnetic fields, so that the magnetization process can be modeled.
\par Our approach also has its limitations. Effects of thermal fluctuations are not included, such that only zero-temperature behavior is analyzed, while at zero temperature effects of quantum fluctuations are largely neglected. On the other hand, both problems could be solved by a numerical treatment of the spin Hamiltonian. Therefore, the approaches based on spin and electronic Hamiltonians are complimentary. Together they can deliver complete microscopic picture of a magnetic material that has now been achieved for kagome francisites. 

\par \emph{Acknowledgements}. We are grateful to Ioannis Rousochatzakis for his fruitful comments about the manuscript. The work of SAN, VVM and VGM was supported by the grant program of the Russian Science Foundation 14--12--00306. AT acknowledges financial support by the Federal Ministry for Education and Research through the Sofja Kovalevskaya Award of Alexander von Humboldt Foundation.

\end{document}